\documentclass[12pt]{iopart}

\usepackage{graphicx}
\begin{document}

\title[]{MEDUSA -- New Model of Internet Topology Using $k$-shell Decomposition}

\author{Shai Carmi$^1$, Shlomo Havlin$^1$, Scott Kirkpatrick$^2$, Yuval Shavitt$^3$, Eran Shir$^3$} 
\address{$^{1~}$Physics Department, Bar Ilan University, Israel}
\address{$^2~$School of Engineering and Computer Science, Hebrew University of Jerusalem, Israel}
\address{$^{3~}$Electrical Engineering Department, Tel-Aviv University, Israel}

\ead {$^1$scarmi@shoshi.ph.biu.ac.il}
\ead{$^2$havlin@ophir.ph.biu.ac.il} \ead{$^3$shire@eng.tau.ac.il}
\ead{$^4$shavitt@eng.tau.ac.il} \ead{$^5$kirk@cs.huji.ac.il}

%\author{Shai Carmi}
%\address{Physics Department, Bar Ilan University, Israel}
%\ead {scarmi@shoshi.ph.biu.ac.il}
%
%\author{Shlomo Havlin}
%\address{Physics Department, Bar Ilan University, Israel}
%\ead{havlin@ophir.ph.biu.ac.il}
%
%\author{Eran Shir}
%\address{Electrical Engineering Department, Tel-Aviv University,
%Israel} \ead{shire@eng.tau.ac.il}
%
%\author{Yuval Shavitt}
%\address{Electrical Engineering Department, Tel-Aviv University,
%Israel} \ead{shavitt@eng.tau.ac.il}
%
%\author{Scott Kirkpatrick}
%\address{School of Engineering, Hebrew University of Jerusalem, Israel}
%\ead{kirk@cs.huji.ac.il}

\begin{abstract}
The $k$-shell decomposition of a random graph provides a different and 
more insightful separation of the roles of the different nodes in such a 
graph than does the usual analysis in terms of node degrees.  We develop 
this approach in order to analyze the Internet's structure at a coarse level, 
that of the "Autonomous Systems" or ASes, the subnetworks out of 
which the Internet is assembled.  We employ new data from  
DIMES (see http://www.netdimes.org), a distributed agent-based
mapping effort which at present has attracted over 3800 volunteers
running more than 7300 DIMES clients in over 85 countries.  We combine this
data with AS graph information available from the RouteViews
project at the University of Oregon, and have obtained 
an Internet map with far more detail than any previous effort. 

The data suggests a new picture of the AS-graph structure, which
distinguishes a relatively large, redundantly connected core of
nearly 100 ASes and two components  that flow data in and out from
this core.  One component is fractally interconnected through peer
links;  the second makes direct connections to the core only.  The model
which results has superficial similarities with
and important differences from the "Jellyfish" structure 
proposed by Tauro et al.\, so we call it a "Medusa."  We
plan to use this picture as a framework for measuring and
extrapolating changes in the Internet's physical structure. Our
$k$-shell analysis may also be relevant for estimating the function
of nodes in the "scale-free" graphs extracted from other
naturally-occurring processes.

\end{abstract}

%Uncomment for PACS numbers title message
%\pacs{00.00, 20.00, 42.10}
% Keywords required only for MST, PB, PMB, PM, JOA, JOB?
%\vspace{2pc}
%\noindent{\it Keywords}: Article preparation, IOP journals
% Uncomment for Submitted to journal title message
%\submitto{\JPA}
% Comment out if separate title page not required
\maketitle

\section{Introduction}
Recent advances in the technology of mapping the topography of the
Internet have made available coarse-grained maps of unprecedented
detail.  In these maps, the nodes are the independently managed
subnetworks (called "autonomous systems" or ASes) which make up
the Internet and the links are direct connections observed between
the ASes.  The physical Internet in 2005 involved over a million
routers, but consists of only about 20,000 ASes.  Each of them is 
assigned a known range of IP address space, 
and thus can be identified from probes, such as Traceroute, which 
report IP addresses. This is an
operationally important coarse-graining, since routing across
multiple ASes is carried out by a different protocol (the "border
gateway protocol" or BGP) than is used inside each AS.  BGP
routing decisions are strongly influenced by business
considerations, many of which involve private information and
cannot be deduced accurately from the Internet's topology.  The
apparent inefficiencies that result -- messages are channeled through
a small fraction of the Internet's links, leading to highly non-uniform
traffic loading -- are a current subject of debate within
the networking community.  It is not clear whether new approaches to long-range data 
routing could be introduced to smooth out the traffic load and, if that were 
accomplished, whether it would lead to increased capacity and shorter delays.  
We hope, in this report, to bring new information and new analytic 
techniques to the debate. 

\subsection{$k$-pruning: Definitions}
We shall show that an informative functional
decomposition of the AS graph can be obtained by the technique of
$k$-pruning, which proceeds as follows:

First, we locate and remove each node with only one neighbor,
removing the link to that neighbor along with the node.  Nodes
removed in this way make up what we shall call the 1-shell.
As this pruning proceeds, we may expose further nodes with one
neighbor (or fewer), for example those which lie in chains or
trees external to the rest of the graph.  These and their links
are removed from the graph and placed in the 1-shell. At the
conclusion of this process we have separated the graph into its
1-shell and its 2-core, the subgraph with only nodes having 2 or
more links to nodes with equal or better connectivity.  The
separation is unique.

We now repeat the pruning process in steps characterized by an
index $k$.  When $k = 2$, we remove all nodes with 2
neighbors remaining from the 2-core, placing them in the 2-shell.  
The process continues, eliminating any nodes reduced to a degree of 2 
(or fewer) by this pruning, until all nodes remaining have 3 or more
neighbors.  This leaves us with the 3-core.  The process is repeated 
to identify the 3-shell and 4-core, and so on. We define a third
subgraph family which will be of interest, the $k$-crust.  This is the
union of the nodes in the 1 through $k$ shells, and the links that
join them. The $k-1$ crust is the complement of the $k$-core.

The process stops at the point when no further nodes remain. The
last nonempty $k$-core provides a very robust and natural definition of
the heart or nucleus of any communications network. We shall
explore this idea below after describing our data set, and compare
it with previous attempts to identify the nucleus of the network
from connectivity information alone. We introduce one further
decomposition.  When we consider the connectivity of the various
crusts, we discover a classic percolation transition \cite{percbook}\cite{Flory}.
For small $k$, the crusts consist of many small clusters of connected
sites.  For sufficiently large $k$, the largest
connected cluster of a $k$-crust consists of a significant fraction 
of the whole $k$-crust, while no
smaller cluster contains more than a few nodes.  The change occurs 
at a well-defined threshold value of $k$.  There is
a significant fraction of the nodes within each large-$k$ crust
which is not part of its largest cluster, and remain isolated.  
Thus we propose to decompose the AS graph (or
any similar scale-free network) into three distinct components:

\begin{enumerate}

\item its nucleus (the innermost $k$-core)
\item the giant connected component of the last
crust, in which only the nucleus is left out
\item the isolated components of the last crust, nodes forming many small clusters.  
These connect to the connected component  
of the last crust only through the nucleus

\end{enumerate}

We shall see that these three classes of nodes are quite different
in their functional role within the Internet. The nucleus plays a
critical role in BGP routing, since its nodes lie on a large
fraction of the paths that connect different ASes.  We find that
it allows redundancy in path construction, which gives
immunity to multiple points of failure.  The connected component of the large-$k$ crusts
could be an effective substrate on which to develop additional
routing capacity, for messages that do not need to circle the
globe.  To achieve this, it may be necessary to develop a pricing
scheme that will reward BGP for avoiding the Internet core
whenever possible.  Finally, the isolated nodes and isolated groups of nodes 
in the last crust  
essentially leave all routing up to the nodes in the nucleus of the network.
Because all their message traffic passes through the nucleus, even when the destination is 
relatively close by, they may be contributing 
unnecessary load to the most heavily used portions of the Internet.  
We shall be monitoring the relative size of this component as a key indicator 
of the evolution of the topography of the Internet.

\subsection{Some History and Previous Work}

We sought in the graph theory literature for a construct that
would characterize redundancy of communications solutions, e.g.,
finding paths between points that need to communicate.  Both
$k$-pruning and $k$-connectivity come to mind, the first being in a
sense an approximation to the second.

$k$-connectivity is a characteristic that captures degrees of
robustness in any parallel distributed system. Two nodes in a
graph are $k$-connected if one can construct $k$ distinct paths
joining them, where "distinct" means that no node and no edge are
used more than once.  A maximal subgraph in which all pairs of
nodes are $k$-connected is thus a generalization of the idea
of a clique, a subgraph in which all nodes are directly connected.
$k$-connected subgraphs were discussed by R Tarjan and J Hopcroft in
the 1970s \cite{HT73}. They showed that an augmented depth-first
search could extract all 2-connected components of an undirected
graph in linear time.  A much more complicated but still linear
time version of this algorithm \cite{3conn} would give the
3-connected components, but it does not generalize to higher $k$. 
In the physics of
conduction in inhomogeneous metals and superconductors, the
2-connected "backbone" of a network modelling such a material is
critical because this is where currents flow in the bulk of the material, 
away from any contacts.  For pictures of such backbones in a 2-D
model system, see \cite{KS85} and for earlier uses of these ideas,
see \cite{K78,K79}.

Bollobas \cite{B84} introduced the $k$-core as a necessary but not
sufficient condition for a subgraph to be $k$-colorable, and proved
under very restrictive conditions that non-vanishing $k$-cores of a
random graph in which all nodes have a Poisson
distribution of the number of their neighbors (this is called an
Erdos-Renyi or \mbox{"E-R"} graph for good historical reasons \cite{ER60}) 
are $k$-connected.  The restrictions were later removed by Luczak
\cite{L91}, who showed that any E-R $k$-core is $k$-connected with
probability approaching unity in the limit of large graphs. He
showed that the probability of finding a separating set of $k-1$
nodes that will break up a $k$-core in random graphs which are
sufficiently well connected to have such a $k$-core in the first place 
vanishes as the number of nodes goes to infinity.

Further study of the $k$-core was triggered by the realization that
$k$-pruning has a sharp threshold when one observes the size of the 
$k$-core as the connectivity is increased in large random E-R
style graphs, and that the $k$-core, when it first appears, occupies a 
finite fraction of the nodes.  
This first order transition was characterized rigorously in \cite{PSW96}.

The extension of this idea to "scale-free" graphs, with
a long-tailed power law distribution of node degree, is a natural one. See
\cite{SKS02} for a detailed discussion of the statistics of the
$k$-cores in preferential attachment \cite{Barabasi} graphs of $10^6$ nodes and
various degree distribution power laws, and for heuristic
arguments linking the power law characterizing the sizes of the
successive $k$-cores to the degree distribution power law in the
original graph.  Several groups have employed $k$-cores and $k$-shells
as an aid to visualizing large networks of naturally occurring
data with power law degree distributions.  See \cite{Bat99,Bat02}
and the work of the Konstanz group \cite{Bau03,G04}.

Finally, after we first reported this work \cite{Trieste}, we became aware of a
concurrently evolving manuscript by Alvarez-Hamelin, Dall'Asta, Barrat and
Vespignani \cite{Al05}.  They also use $k$-core
decomposition on data for the AS-graph, and identify
$k$-shells.  They emphasise in particular the similarity between the
degree distribution within the $k$-cores and that of the
original graph. Their $k$-shell visualization program can be seen on the web
at \cite{lanet}.

There have been several previous attempts to characterize the core
of the Internet AS graph.  Subramanian et al. \cite{SARK02}
identified a group of ASes that have no declared provider (this
requires side information, cannot be deduced from the graph itself) and used heuristics to find the
largest loose clique, a group of N nodes such that each is connected to
at least N/2 of them and no two nodes are more than two hops away
from each other.  A similar
approach was suggested by Ge et al. \cite{GFJG01}. Tauro et al.
\cite{TPSF01} define the core as the maximal clique that contains
the most connected node.  Unfortunately, this proves very sensitive to the addition
or absence of a link.  Bar et al. \cite{BGW04} suggested to look
for loose cliques, but unlike \cite{SARK02,GFJG01} used the graph
topology without side information.  They used the dense
$k$-subgraph heuristic of Feige et al. \cite{FKP01} to find for each
size $l$ the densest subgraph of this size.  The result is a series
of clusters, one for each cluster size.  It is not clear (on the data
they analyzed) which size is the most appropriate to choose or how
robust the solutions are to changes in the details of the search
heuristic.  Our definition of a core is more robust and simpler to
calculate than any of the above solutions.

\subsection{Data and methods used}

The data we analyze is gathered from a constellation that currently (YE2005)
has more than 7300 software agents, each running on a different computer.  
The agents have been downloaded from the DIMES website ( http://www.netdimes.org ), 
and run on PCs in more than 85 countries around the world.  Upon initialization, 
each agent requests, from a central server, a package of measurements to be performed. 
Although a broad class of measurements can be assigned to the agents, at 
present most packages consist of a mixture of "traceroute" and "ping" commands 
directed at a random sampling of more than 5,000,000 destinations in the Internet's 
IP-address space, chosen to cover all prefixes known to have been assigned.  Each agent 
receives a different measurement package to execute, and performs up to four 
such measurements each minute, returning the results to the central server and 
getting more work to do as each package is completed. The portfolio of randomly 
generated packages being dispatched to the agents is constantly evolving.  
Some address random distant destinations.  Others attempt to keep the 
destinations close to the observed location of the agent, in order to 
discover peering connections, which are often not publically disclosed.  
Since we seek complete coverage but do not require that it be uniform, 
we integrate all the measurements to obtain a graph which describes all nodes and links observed to carry packets during a prescribed period of time. We treat each link discovered as undirected. For a more detailed description of the objectives and issues that guide the DIMES design, and some of the additional things that DIMES can do, see the technical report \cite{SS05}.

For purposes of this study, all addresses which can be resolved to lie within a single AS are considered to be a single node.  Two successive addresses discovered to lie within different ASes define a link between those ASes. All incoming raw data are archived to permit reexamination as new questions arise. AS resolution was done with the aid of prefix tables which were refreshed each week. The information resolved to the AS level is stored as edges and links in a database, along with information such as the time when each edge or link was discovered, by whom and in the course of which measurement, 
and the time when and by whom each was most recently observed, or "validated".  There are rare situations in which the traceroute data may be corrupted, so, in our current study, we considered only links that had been seen on at least two different occasions.  
This is conservative, as we are probably discarding some valid links. 
% During the three months considered in this work, new ASes were discovered continuously, because they constantly are added, but mostly because of the increasing effectiveness of our measurement constellations.  We can more reliably measure ASes which disappear during the study.  We observed that xx ASes that were seen up to March 1, 2005 were no seen after April 1.  In addition yy of the ASes known to exist on March 1 disappeared during April, and an additional zz ASes vanished during May 2005.   

Table \ref{Table_I} shows the number of nodes and links discovered by the DIMES searches during the period from March 1, 2005 until June 1, 2005.  The number of users and agents participating in the DIMES effort was increasing steadily during this period.  On March 1, DIMES had 700 users deploying 786 agents, and by June 1 that number had increased to 2145 users and 3069 agents.   Most of the increase occurred in mid May, following a news story about DIMES which appeared in Science \cite{Science_news} 	 and subsequent discussion of the project at slashdot.com.  During the period from March 1 through mid-May, DIMES accumulated roughly 30 million measurements.  In the second half of May, another 30 million measurements were collected.  As a result, we have tabulated the information found for three periods, March 1 until June 1, 2005 ("03-06"), April 1 to June 1 ("04-06"), and May only ("05-06"). In our figures, we will use the 03-06 data set unless the caption further specifies that a comparison between the different intervals is intended.

Also in Table \ref{Table_I} are the numbers of nodes and links found by using the University of Oregon's RouteViews \cite{RouteViews} project database, which tabulates the preferred border gateway protocol (BGP) routes broadcast to the rest of the Internet by 85 of the largest ASes, which peer with the Routeview project's eight routers from 200 locations around the world.  We used for the BGP data sets only the nodes and links reported during each period.  One of the intriguing issues in understanding the Internet at this coarse level is the differences between the DIMES and BGP views.  In this period, BGP tables identify just over 20,000 different ASes that are publically accessible, while DIMES was able to see just over 14,000 ASes (including a few hundred that BGP had not mentioned). The differences between the links observed by the two methods is even greater.  Over the full three month period, DIMES and BGP both saw about 28000 links, while BGP reported 21000 links that DIMES did not observe, and DIMES found 15000 links not disclosed by BGP. The Internet graph that results from the union of the two observations is the densest graph reported to date, with an average of just over 6 neighbors for each node.    Data for the entire three month period is available on the web at \cite{DIMESURL}.

\begin{table}[thb!]
\center{
\begin{tabular}{|c||c|c||c|c||c|c|}  \hline
{ \bf data } & {\bf DIMES } & {\bf DIMES } & {\bf BGP } &  {\bf BGP } & {\bf all } & {\bf all }\\
{ \bf set } & {\bf edges } & {\bf nodes } & {\bf edges} &  {\bf nodes } & {\bf edges} & {\bf nodes}\\
\hline
03-06 & 42174 & 14386 & 48141 & 20340 & 62920 & 20556  \\
04-06 & 40410 & 14274 & 46689 & 20206 & 60754 & 20463  \\
05-06 & 37201 & 13990 & 44908 & 20084 & 57516 & 20356  \\
\hline
\end{tabular}
\caption{Number of ASes and AS-AS links found in the DIMES measurements, Routeviews BGP tables, and their union.  Top line: March to June data; mid line April to June; bottom line: May only}
\label{Table_I} }
\end{table}

\section{Analysis of the AS graph data}

Most discussion of the role of ASes has been expressed
in terms of their degree, $z$ -- the number of links to other ASes.  In \cite{SS05} the 
degree distribution for this data is plotted, giving the result seen previously, that for a 
significant range of $z$, the probability of a node having degree $z$, 
$p(z) \propto z^{-\gamma}$ where $\gamma =$ roughly 2.2.  More subtle correlations have also been studied
as a function of degree. We shall consider some of the more instructive analyses in the literature, using the $k$-shell as our principal index to each AS's behavior on this data, which is more extensive than that found in previous publications. 

\subsection{$k$-shell, $k$-crust, and $k$-core characteristics}
    
\begin{figure}[htb] \centering
\includegraphics[totalheight=3.5in]{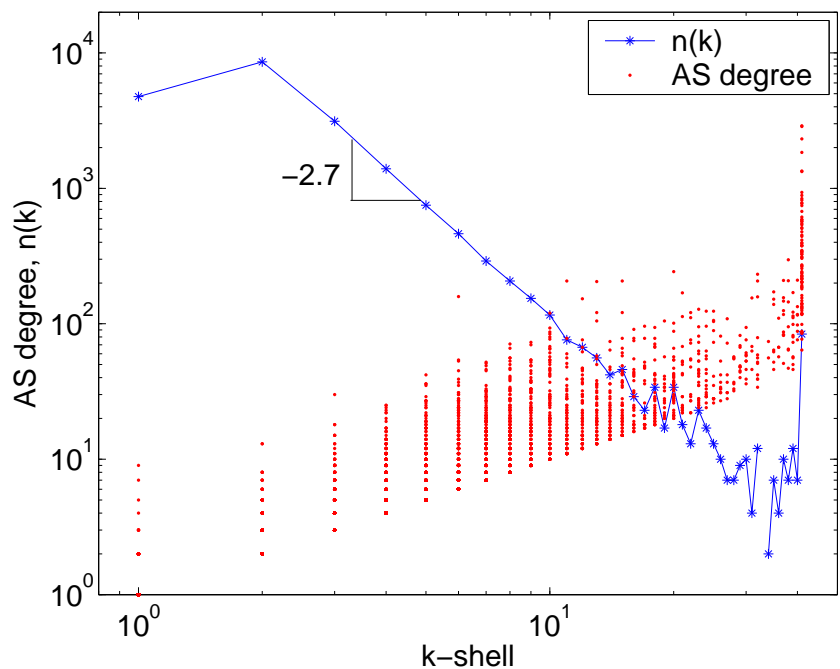}
\caption{\small{Scatter plot of node degree as a function of $k$-shell number.  Also shown is $n(k)$, the size of each $k$-shell, which decreases roughly $\propto k^{-2.7}$.}} \label{Shells}
\end{figure}

Fig \ref{Shells} shows that $n(k)$, the size of each $k$-shell, decreases with a power law distribution, 
$n(k) \propto k^{-\delta}$,
where the exponent $\delta$ is about 2.7.  The first shell is an exception to this rule, most likely because ASes, even at this level, are
incented to purchase additional connections in order to ensure
reliable service to their customers. We expect that the tendency to provide multiple links is increasing, and that this shell will decrease further in the future.  In \cite{SKS02}, it is argued that $\delta \ge \gamma$, and using a particular random preferential attachment model, they conclude that a $\gamma$ of 2.2 leads to a $\delta$ value of 2.6, so our observations are consistent with these random scale-free results.  Of course that also implies that
from these power law exponents alone,  little can be inferred about any detailed structure or subtle correlations in the actual AS graph.

From the scatter plots in Fig \ref{Shells}, we see that there is a significant variation of the degrees in each $k$-shell, within a general trend that the larger $k$-shells contain ASes with higher degree.  
Each $k$-shell contains nodes with a roughly ten-fold range of degrees.  It may seem surprising that the range of almost 10 even applies to the 1-shell, where there are 3 sites (out of 5000) that have 5, 7 and 9 neighbors, but inspection of the data shows that these in fact collect and concentrate data from stars with 4 and 8 ASes, and a tree of 6 ASes,  and pass it along to a single provider deeper in the network. The variation in degree persists until, at roughly $k=30,$ the power law decrease in the sizes of the shells stops.  There is then, from $k=30$ to $k=40,$ a region in which the sizes of the $k$-shells are rather small.  Finally, the nucleus is observed to be the $k=41$-shell in this data set, and it consists of some 84 nodes, ranging in degree from about 100 to over 2000.  This basic behavior has already been observed in the $k$-shells of random graphs grown with a preferential attachment mechanism tuned to produce the same power law decrease in the probability distribution of the node degrees \cite{SKS02}.  However, the router level distribution obtained in the DIMES measurements shows a bigger gap between the end of the power law decrease in $k$-shell sizes and the nucleus \cite{SS05}.  The largest degree seen for nodes in the nucleus of the router level network is somewhat less than that seen for the last shells in the inflow regime of that network. This is most likely a result of the characteristics of actual hardware. 

\subsection{$k$-connectivity}

\begin{figure}[htb] \centering
\includegraphics[totalheight=3.5in]{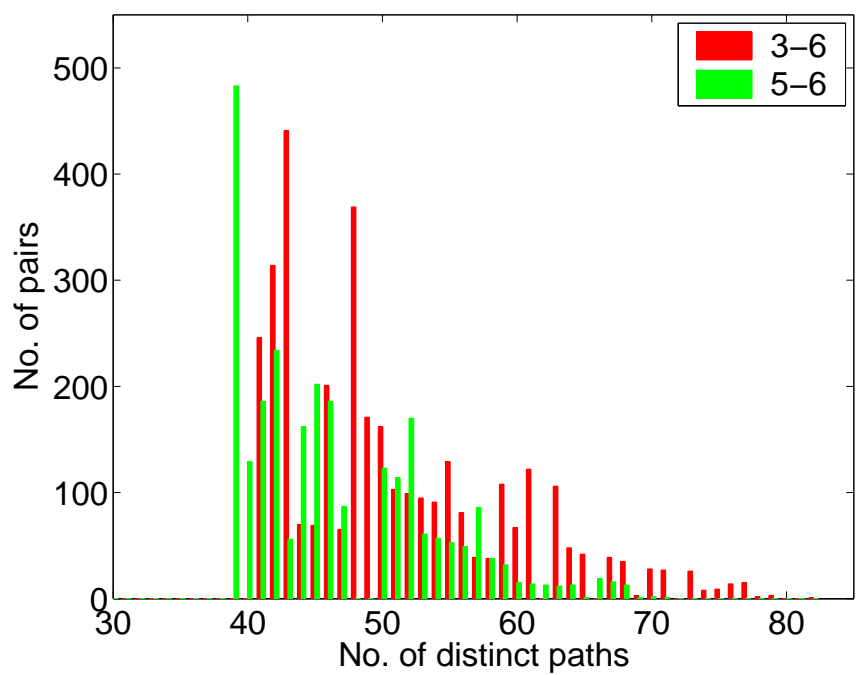}
\caption{\small{Distribution of number of distinct paths possible between any pair of nodes, observed in the nuclei of the 03-06 and 05-06 AS graph data sets.  For all such pairs, the number of possible paths equals or exceeds the $k$-core index, which is 41 for the 03-06 nucleus and 39 for the 05-06 nucleus.}} \label{Coreflows}
\end{figure}

We test for the degree of connectivity of the nodes in the AS network by the standard method of calculating
the maximum flow possible between pairs of nodes using each intermediate link and node only once.  
The number of possible node and link distinct paths between a given pair of nodes is this maximum flow.  
It is relatively easy to calculate, using a max-flow algorithm \cite{Flow_reference} with unit link capacity.  The constraint that each node is used only once is imposed by putting a fictitious link with unit capacity between incoming and outgoing flows 
at each node.
In Fig. \ref{Coreflows}, we show the distribution of flows between all pairs of nodes in the nucleus of our data set, for the periods 03-06 and 05-06.  The nucleus occurred at $k=41$ for the longer period and contained 84 nodes.  For the last month only, the nucleus included 73 nodes and occurred at $k=39$.    In each case, the smallest number of paths observed was equal to the $k$-core index of the nucleus.
The probability distribution peaks, in each case, at or near the minimum value, $k$.

\begin{figure}[htb] \centering
\includegraphics[totalheight=3.5in]{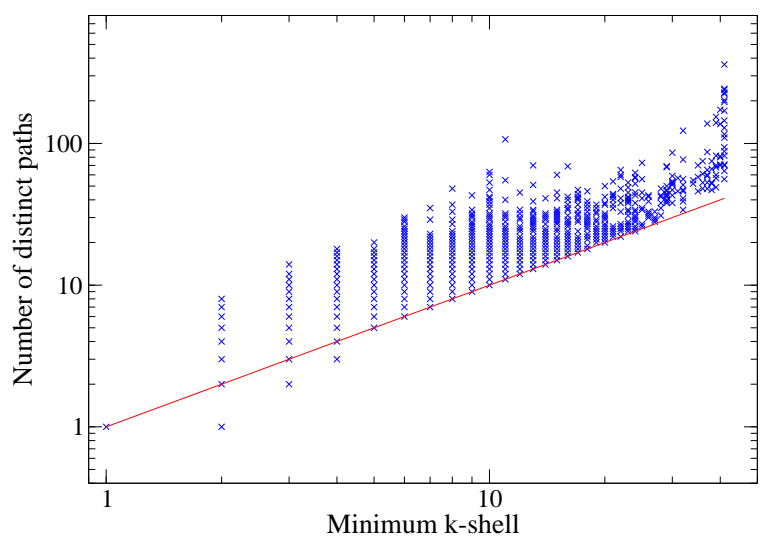}
\caption{\small{Scatter plot of the number of distinct paths possible between any pair from a sample of 50,000 in the connected part of the AS graph, plotted against smaller of the two nodes' $k$-shell indices. }} 
\label{Exceptional_Flows}
\end{figure}

We also evaluated the number of distinct paths between pairs of ASes at arbitrary locations in the Internet.  In almost all cases, the smaller of the $k$-shell indexes of the two endpoints provided a lower bound to the number of distinct paths between the pair, as shown in Figure \ref{Exceptional_Flows}.  Violations of this "rule," which is not a rule since the AS graph is hardly random, occur for $k=2$, 3 and 4.  Examining the nodes for which exceptions are seen, we find that these form small clusters of two or three nodes in the same $k$-shell which link to each other and which then, as a group, link only to less than k nodes deeper into the Internet.  This is a natural result of having multiple ASes under common management, as well as the lower cost of links which are 
local to a given region.

\begin{figure}[htb] \centering
\includegraphics[totalheight=3.5in]{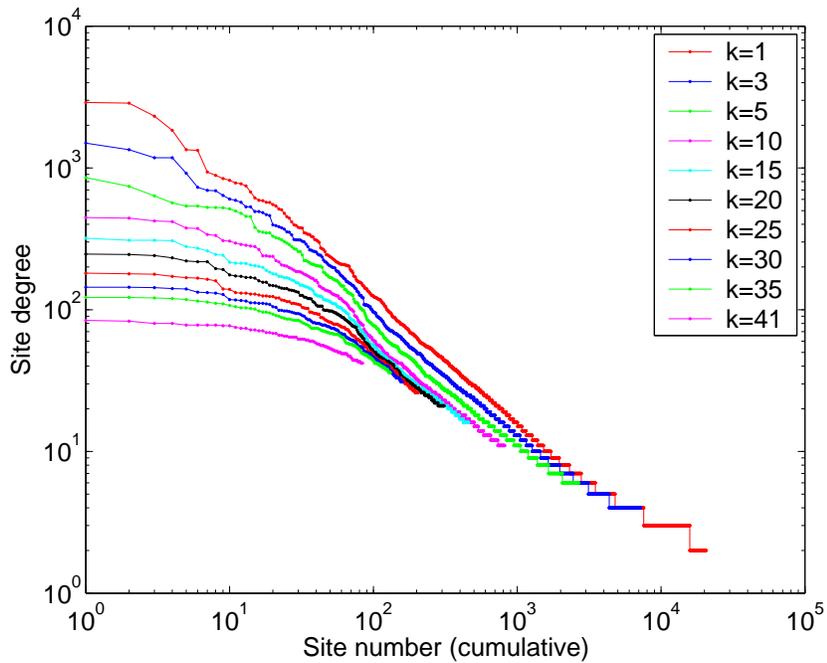}
\caption{\small{Zipf plot of the node degrees (against their sequence number when sorted into descending order, generating a cumulative distribution) in the original AS graph and in each $k$-core.}} 
\label{kcore_degrees}
\end{figure}

A second question is whether we can characterize in some simple way the structure of each successive $k$-core.  If the original graph is randomly connected, with only its degree distribution specified \cite{Molloy_Reed}, we expect that each stage of pruning to excise a $k$-shell will remove all nodes with less than k neighbors, and reduce the number of nodes remaining with degree greater than k in proportion to their degree.  This means that the high degree tail will be trimmed and the power law distribution remaining will become slightly steeper.  Some sort of crossover, whose nature requires a more detailed calculation, will occur as the cores become smaller.  Figure \ref{kcore_degrees} shows this process occurring for our AS graph.  Between $k=30$ and the nucleus ($k=41$), the $k$-core appears to be some sort of dense random graph, and no longer exhibits any power law regime in its degree distribution.  

Both Shalitt et al.\ \cite{SKS02} and Alvarez-Hamelin et al.\ \cite{Al05} have observed that the beginning of this process can be fairly well described by treating the early $k$-cores as pruned, scaled-down copies of the original AS graph.  From the arguments above, we see that this is an approximation, and does not give insight into the crossover that occurs or the nature of the deeper $k$-cores.  In fact, for our three data sets, the nuclei are densely populated random graphs, with .70 of the possible links present in the 03-06 and 04-06 nuclei, and .72 of the possible links present in the slightly smaller 05-06 nucleus.

\section{The Medusa model for the AS graph}

\begin{figure}[htb] \centering
\includegraphics[totalheight=3.5in]{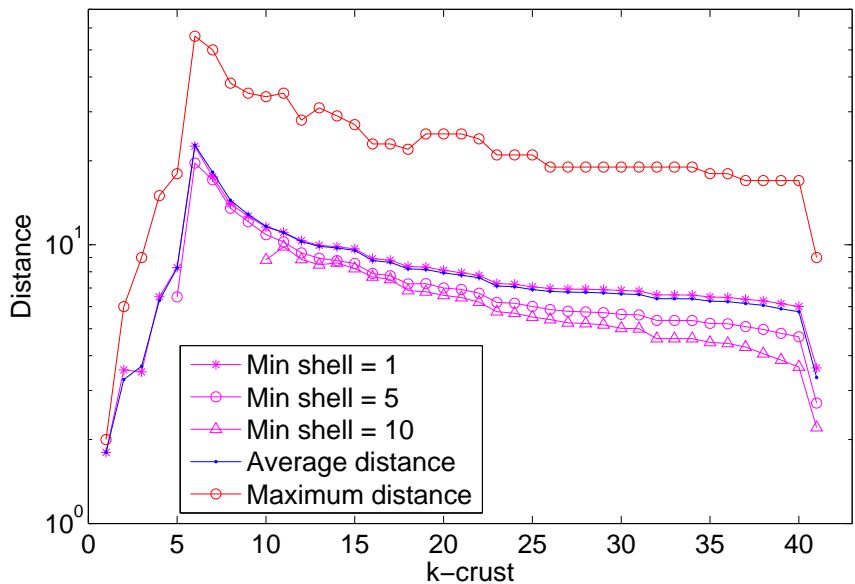}
\caption{\small{Average distance between nodes in each crust, given by the solid line.  Average separations between subsets of the crusts in which the minimum k-shell index of either source or destination is 1, 5 or 10 is indicated with data points and dashed lines.  Also the diameter of each crust (maximum distance between any pair of nodes) is shown above the averages.  The peak at $k = 6$ is a clear sign of criticality in the formation of an infinite percolation cluster.}} \label{distances}
\end{figure}

\begin{figure}[htb] \centering
\includegraphics[totalheight=3.5in]{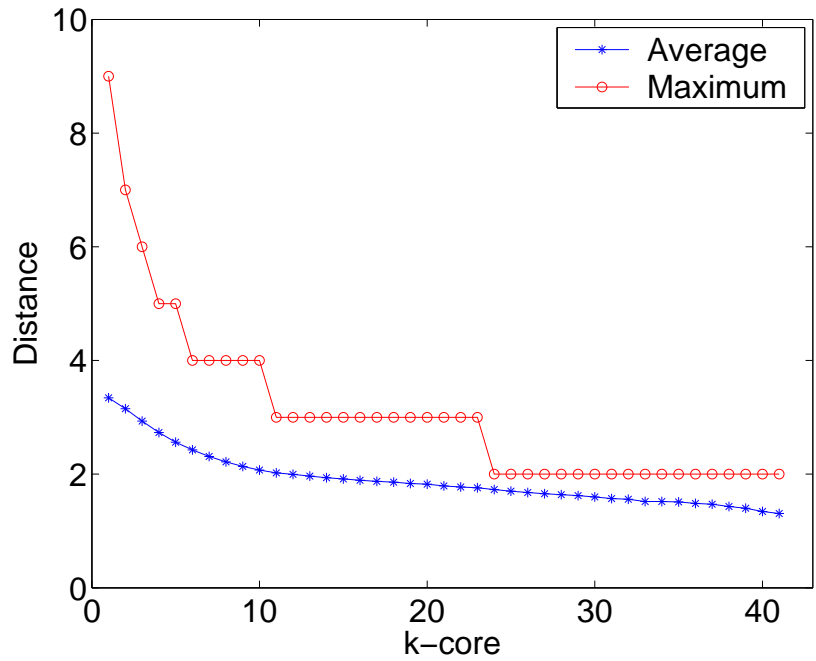}
\caption{\small{Average distance and maximum distance (diameter) between nodes in each $k$-core.}} 
\label{coredist}
\end{figure}

Since the fraction of ASes which are in the ultimate core of the Internet, the nucleus, is less than one per cent, we have studied the capabilities of the $k$-crusts remaining when the nucleus is removed, in order to see what its role is in BGP routing of data, or could be under alternative schemes.  We use simple graph-theoretical constructions for this analysis.  First, in Figure \ref{distances}, we find that 
the diameters (the maximum distance between any two nodes) of the crusts are
much greater than the diameter of the whole graph.  The same is true for the average distance in hops between nodes in crusts.  For the first few crusts the average distance (which is computed only between nodes that can reach each other using links within the crust) starts rather small, because the crusts are not connected, but consist of many isolated clusters of nodes.  By the time we reach crust 6, however, the average distance is more than 20 hops and the diameter more than 60.  These distances decrease steadily as the crusts reach deeper into the Internet.  In the 40-crust, the average distance is only 6 hops and the diameter slightly less than 20. Both figures are roughly twice the values for the entire graph, whose diameter is 9 hops and whose average node-node distance roughly 3.4 hops.  The distance between nodes in a crust is also affected by the $k$-shell index of the source node.  In Figure \ref{distances}, we also see that the average distance from a site in the inner $k$-shells to any other site in each crust decreases as the sources are restricted to lie deeper in the crust.  We see the influence of the nucleus even more strongly in Figure \ref{coredist}, which shows diameter and averaged node-to-node distance for each $k$-core.  The diameter remains 2 as $k$ decreases from 41 (the nucleus) to 24, and only increases to 3 as $k$ decreases to 11.  It increases dramatically as the shells with $k < 11$ are then added to the core nodes.

\begin{figure}[htb] \centering
\includegraphics[totalheight=3.5in]{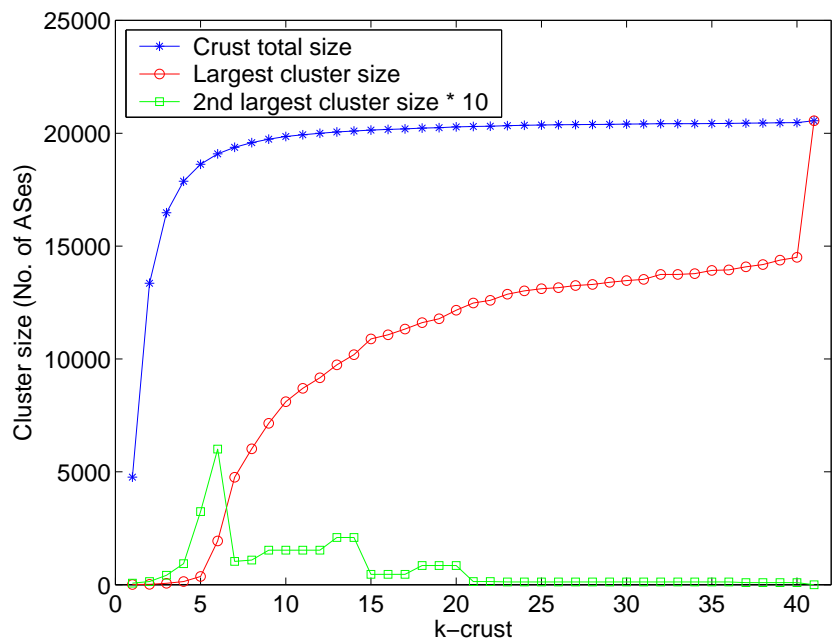}
\caption{\small{Size of the largest and second largest cluster in each $k$-crust.}} \label{Percolation}
\end{figure}

Analyzing this more carefully by testing for an "infinite cluster," in Figure \ref{Percolation}, we see that a percolation threshold occurs as we increase the depth of the crust.  For $k < 5$, the largest cluster of connected ASes is extremely small, but it rises sharply both in absolute number and as a fraction of the total number of nodes in the crust for $k>5$. This has the typical characteristics of a percolation threshold \cite{percbook}, in that this one cluster dominates above $k = 6$.  The second largest cluster is also plotted under the same conditions (magnified to make it more visible) in Figure \ref{Percolation}.  It is significant only very close to the threshold, at $k = 5$ or $k = 6$. In most percolation phenomena, the "infinite cluster" covers almost the entire graph.  Here, however, the largest cluster in the crusts never occupies all of the crust.  We see that in this data, some 30 per cent of the nodes only connect to the rest of the Internet through their links to the nucleus set of nodes.  The volume of the largest cluster saturates at roughly 70 per cent of the crust nodes.  The percolation threshold that we see as we add shells to thicken the crust of the network resembles the percolation threshold observed by Cohen et al. \cite{CHA02}\cite{CEAH01} in a process they call an "intelligently directed attack" on the Internet, where they remove nodes from the network in descending order by degree, although our process proceeds in the opposite direction.  They reported results for large randomly generated graphs with the degree distribution \cite{Molloy_Reed} characteristic of the AS graph.

\subsection{The nucleus of the Internet}

The nucleus defined by $k$-pruning as the true "core" of the Internet emphasises flexibility of connections rather than the sheer degree of its nodes, their closeness to all other nodes as measured by hop count, or attempting to characterise their bandwidth or link speeds.  The "Jellyfish" construction is the simplest of the previous discussions of constructing such a core \cite{TPSF01}.  This starts with the AS having the highest degree.  In our data this is ATT Worldnet, AS No. 7018, with nearly 3000 connections found 
in our 03-06 data set.  
Next the core is defined as the largest clique which can be constructed by starting with this AS, adding nodes each of which is connected to all those previously added.  This heuristic depends upon the search order in which additional nodes are considered for addition to the clique being formed.  The most obvious way in which to perform the search is in decreasing order by degree.  In this way one finds a clique of 24 members, 11 of whom are US-based carriers well known for their international presence (e.g., Global Crossing, Level 3, Sprint, Qwest, Verio, Metromedia...).  The remaining carriers identified in this way are European -- three in the UK, three in Germany, one each in Italy, the former Soviet Union and Lebanon.  Unfortunately this identification is not particularly robust.  One could also order the search using only the number of connections that each of the sites participating in the innermost $k$-core has with other members of our nucleus, since these are the links that will be involved in the eventual clique.  This causes about 25 per cent of the list to change, affecting mostly the European sites.

Our definition of the core is, first of all, robust.  We compared the composition of the nuclei found in the 03-06, 04-06, and 05-06 data sets (in each case, keeping only those nodes that were observed on two or more occasions).  There were 84 nodes in the 41 core of the 03-06 data set, 82 nodes in the 40 core of the 04-06 data set, and 73 nodes in the 39 core of the 05-06 data set.  The nodes contained in the 05-06 nucleus are all found in the two larger sets, and the nodes in the 04-06 nucleus are all also found in the 03-06 nucleus. Looking from the opposite direction, the two nodes in the 03-06 nucleus which are not in the 04-06 nucleus are only one shell outside of it.  The 11 nodes in the 03-06 nucleus not in the 05-06 nucleus were also located close by, with most lying 1-3 shells outside the nucleus, and the most extreme change being one node which was found 7 shells from the nucleus, in shell $k=32$.  We believe that this definition of the true "tier-1 ISPs" will be useful in future studies of the temporal evolution of the Internet, as its changes over time will be a valid measure of changing roles.

Finally, the nucleus defined as the maximal $k$-core is much more democratic and world-wide than the identifications that have resulted from maximal cliques or loose-cliques.  In addition to the US carriers with extremely high degree, we found 11 more Far Eastern ASes with high degree, two each for China, Japan, Korea and Australia, and one each in Hong Kong, Malaysia and Singapore.  Pakistan has a large AS in the nucleus, but India appears to connect to the rest of the world through smaller ASes or through ASes which are foreign-owned.  South Africa has a single AS in the nucleus.  In Europe, at least 15 countries participate in the nucleus.  The UK has the most diverse presence, with 9 ASes, Germany has 5, Denmark 4, and Spain, Sweden, France and the Netherlands have two ASes each. Thus most countries in eastern and western Europe have an AS participating in the 41 core, where they have the maximum freedom of interconnection.  The remaining ASes in the nucleus are smaller US and Canadian ASes.  Data communications in today's Internet, we find, have become much less monolithic and "backbone-oriented" than is commonly presumed.    

Having identified the nucleus ASes of the Internet, we revisited the question of what sources the Oregon RouteViews project \cite{RouteViews} is interrogating.  The 85 ASes currently (YE 2005) listed on the project website were actually interrogated at over 200 router locations, and they peer with eight routers managed by the Univ. of Oregon.  We identified the k-shell membership of the ASes, and found that half of them lie in our 03-06 nucleus, with about 10 ASes occurring in shells 1-5 and the rest more-or-less uniformly distributed over the shells in between. 

\subsection{Three components}

\begin{figure}[htb] \centering
\includegraphics[totalheight=3.5in]{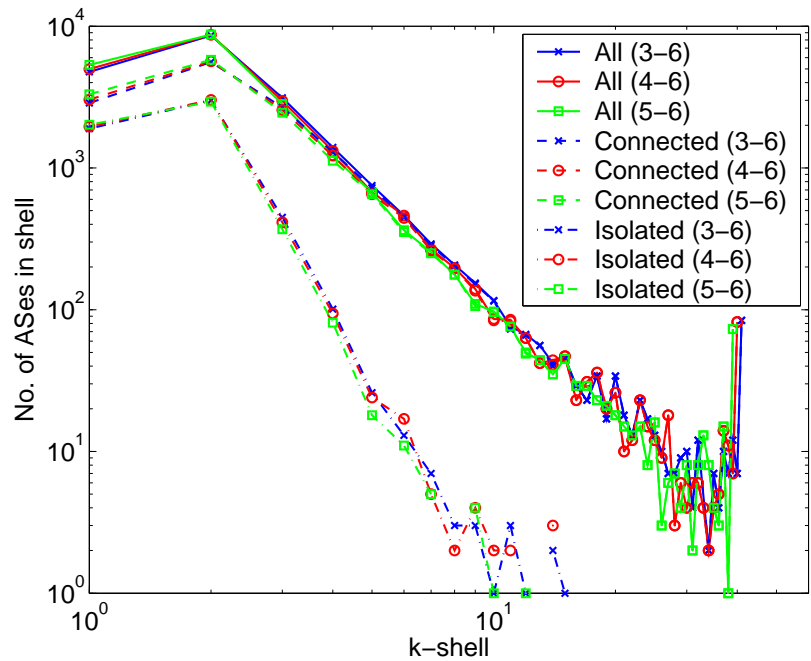}
\caption{\small{Sizes of the three components of the AS-graph, measured over 3, 2, and one month intervals, follow nearly identical curves.  The size of the isolated component decreases roughly as $k^{-5}$.}} \label{Three_components}
\end{figure}

Next we plot the sizes of the three components, the isolated nodes and the connected component, including the nucleus, against their $k$-shell indices, and obtain the power law distributions shown in Figure \ref{Three_components}.  The curves are essentially identical for the three time periods studied, showing that our measurements had stabilized during this period, covering what they were capable of observing within only a few weeks of observation.  Subsequent to the period under study, DIMES has continued to discover additional nodes and links, using further enhancements to the measurement portfolio.

\begin{figure}[htb] \centering
\includegraphics[totalheight=3.5in]{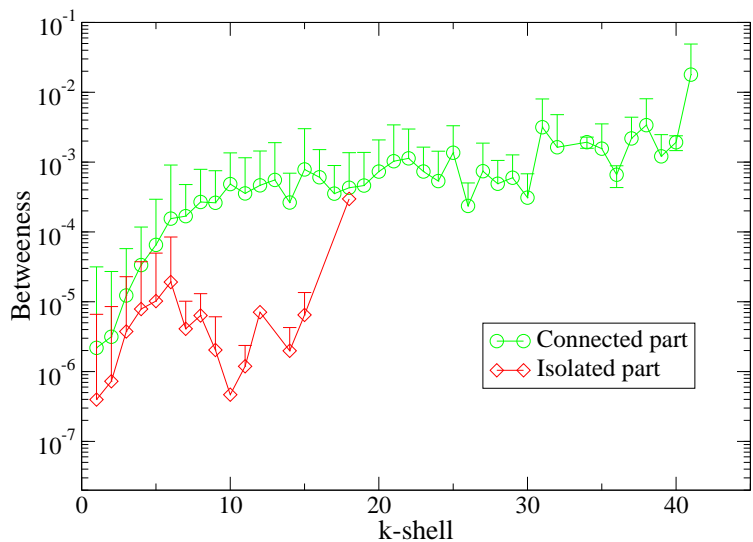}
\caption{\small{Average betweenness versus $k$-shell.  Isolated nodes' values are plotted in red, nodes in the connected component in green.}} \label{Betweenness}
\end{figure}

\begin{figure}[htb] \centering
\includegraphics[totalheight=3.5in]{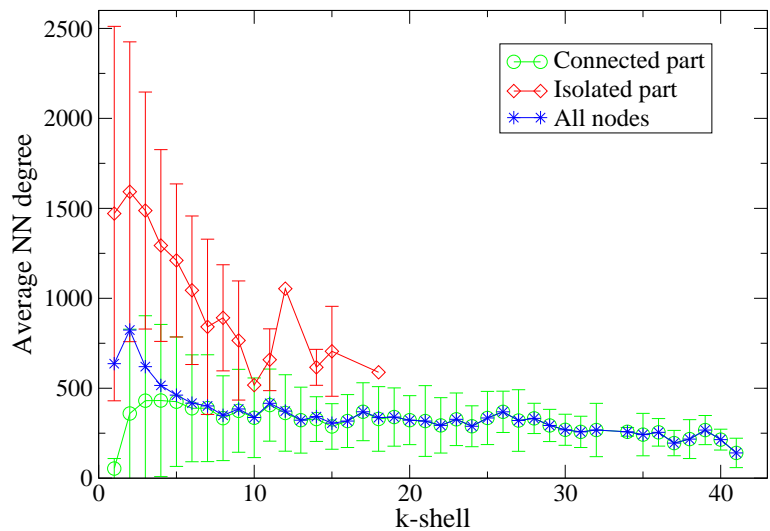}
\caption{\small{Average nearest neighbor degree for nodes in the isolated (red) and connected (green) components of the AS graph. The average over all nodes is shown in blue.}}\label{avg_NN}
\end{figure}

\begin{figure}[htb] \centering
\includegraphics[totalheight=3.5in]{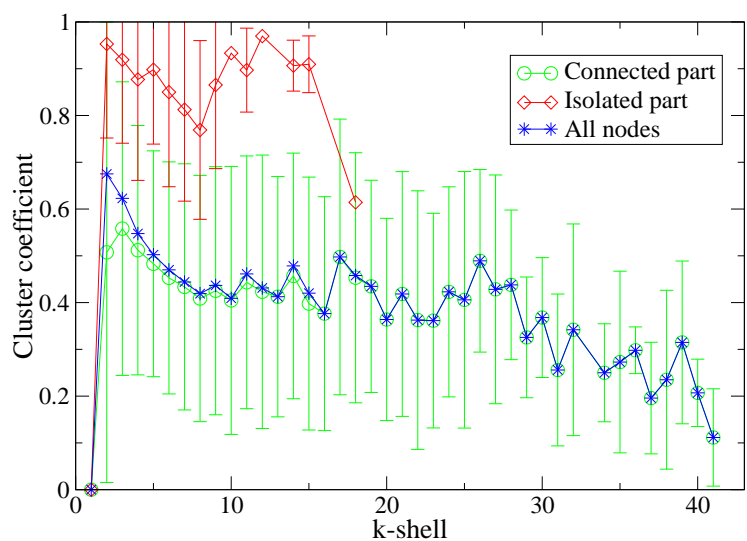}
\caption{\small{Average cluster coefficient for nodes in the isolated (red) and connected (green) components of the AS graph. The average over all nodes is shown in blue.}} \label{avg_CC}
\end{figure}

The separation of the last crust into connected and isolated nodes also signals large differences in the environments of the nodes in the three groups.  To demonstrate this, we plot three of the traditionally studied environmental averages \cite{Vesp_book}, betweenness (in Figure \ref{Betweenness}), average nearest neighbor degree (in Figure \ref{avg_NN}), and cluster coefficient -- the fraction of possible links between immediate neighbors that are realized -- (in Figure \ref{avg_CC}).  For simplicity, we include the characteristics of the 
nodes in the nucleus (the 41-shell) in the same color and style as the data from the connected component of the crusts.

Betweenness is studied as a convenient stand-in for a traffic measurement (which is not available because the information is contained in buffer monitors of routers around the world, and no one organization has the privilege to access this data).  It presumes a traffic model in which every defined point on the Internet sends one packet to every other point on the Internet.  The "betweenness" of a link or node is then the fraction of these packets that pass through that link or node.  This is not a particularly realistic traffic model. It might be more plausible to consider only messages from leaves to leaves, (we have done this, and the results do not differ much, given the high fraction of the nodes which are leaves).  More realistic would be a model in which the number of packets sent varies widely from node to node, but we have no data on which to base this.  So in Figure \ref{Betweenness} we use the conventional traffic model, assume shortest path routing on the AS graph, and show the betweenness of nodes.  Only the nucleus nodes have betweenness of more than a per cent.  The isolated nodes in the last crust, with one exception, and the connected nodes in the last crust have figures of less than $10^{-4}$.  The remaining k-shells of the connected component and the one isolated node in the 18-shell show a steadily increasing value of betweenness, ranging from $10^{-4}$ to about one-tenth of the value shown by the nucleus sites.  The one exception, an isolated node in the 18-shell, is an American hosting facility (a "colocation center") with three ASes in the $k=1$ shell as customers, and direct links to 18 of the most highly connected US and European ISPs in the nucleus.  It thus is a borderline example of letting routing be done by the nucleus, and with this many connections it may in fact also carry peering traffic.

The average nearest neighbor number, as shown in Figure \ref{avg_NN}, and the cluster coefficient, shown in Figure \ref{avg_CC}, are two similar measures of the degree to which the immediate environment of a site has the characteristics of a dense random graph, i.e. of the nucleus.  As a result, the isolated nodes show high values of both quantities, and lie almost entirely outside the range of the smooth trend that the connected nodes exhibit.  The large variances seen, especially in the cluster coefficient graph, are probably a result of the high degree of connectedness that the high k-shells exhibit, and the relatively large fraction of links made from early k-shells to later k-shells, including the nucleus itself.    

\subsection{Fractal structure of the connected component}

\begin{figure}[htb] \centering
\includegraphics[totalheight=3.5in]{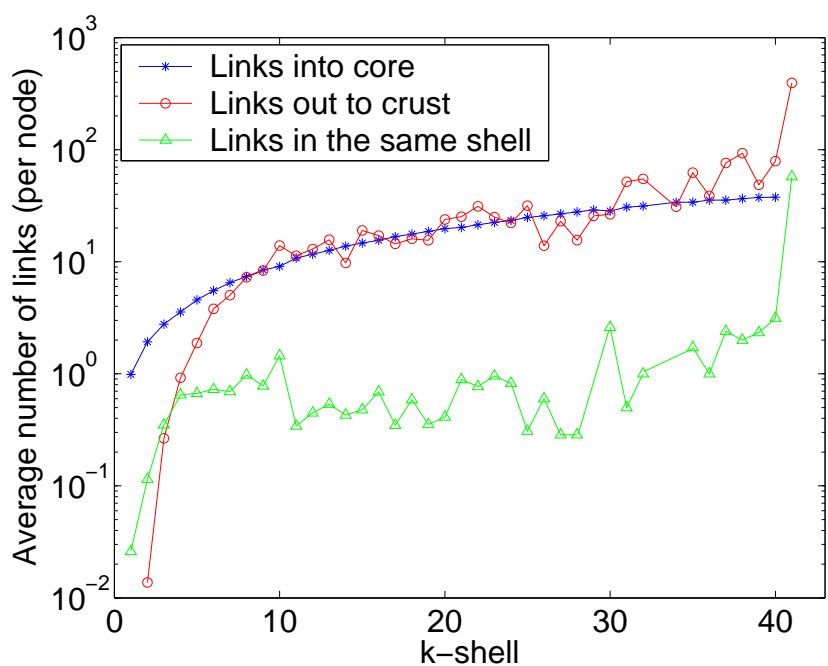}
\caption{\small{For each $k$-shell, we plot the number of links per node that come from nodes further out ($<k$) in the Internet, are made to the same shell, or are made to deeper ($>k$) shells.}} \label{linksinout}
\end{figure}

Finally, we discuss the nature of the connected part of the AS graph. One objective of the DIMES measurements is to make possible more accurate statistical models of the Internet both in its present state and in the future as it evolves. A first insight into the nature of such a model comes from plotting the average number of links each node makes with nodes in deeper shells, in Figure \ref{linksinout}.  This stays close to $k$ for almost all shells, indicating that connected structures within a shell (e.g. the chains and trees exposed in the 1-shell) are only a small correction to the basic structure.  Surprisingly, the number of links per node reaching each shell from shells lying outside also tends asymptotically to $k$, while the number of links made per node within each shell is nearly constant, roughly equal to one. Note that the average number of links out in Figure \ref{linksinout} tracks nicely with the betweenness observed in the connected part of the network, as shown in Figure \ref{Betweenness}, as it should since most messages counted in this measure start or stop in the more populous outer parts of the Internet. 

\begin{figure}[htb] \centering
\includegraphics[totalheight=3.5in]{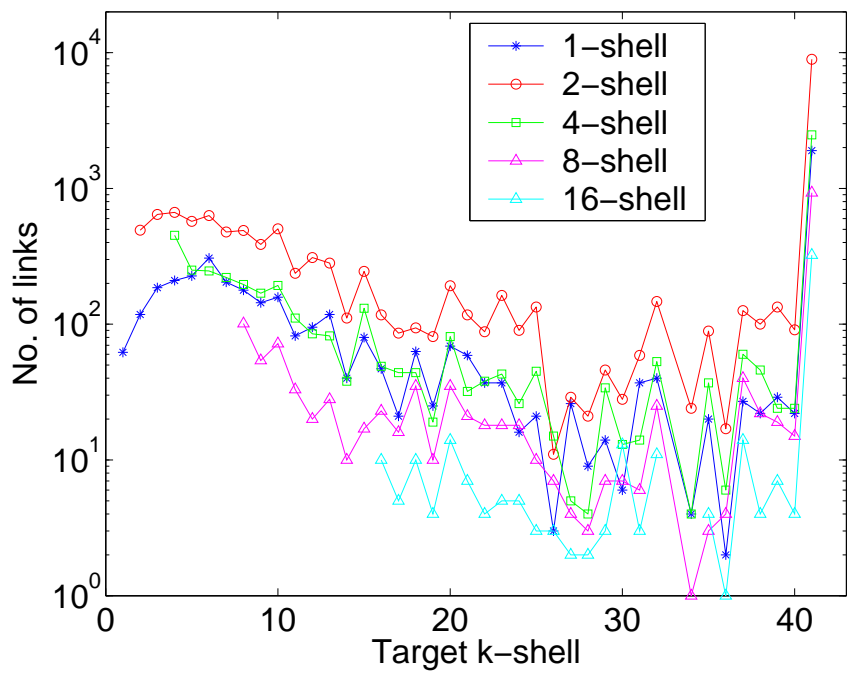}
\caption{\small{For a selection of shells, we plot the number of links that are made to each deeper shell in the Internet.}} \label{linkages}
\end{figure}

Figure \ref{linkages} breaks these averages into results for shell-to-shell linkages.  This shows that the links from each shell extend over a wide range of depths further into the Internet. Seeing a wide extent in range as well as number of linkages makes us suspect that some sort of fractal structure is present.  We analyze this with a specific test below.  

We differ fundamentally in the labelling of nodes outside the nucleus from the "Jellyfish" model \cite{TPSF01}. They label the outer nodes by performing a breadth first search from the core outwards, and identify rings of nodes 1, 2, and more steps away. Our Medusa labelling is from the outside in, using k-shell index, and gives a longer scale.  They consider leaf nodes (with only a single connection) to form tendrils hanging from the core and the mantle of their "Jellyfish," and notice that a surprisingly large fraction of the leaf nodes are connected to the innermost core.  (Our isolated component is only connected to the nucleus -- like the tendrils of a real Medusa -- but our tendrils may consist of clusters as well as individual sites.)  It is not clear which picture will prove most useful as a basis for generating models or analyzing the Internet's capacity.  The Jellyfish's links connect ring $n$ to ring $n-1$, but nodes in adjacent rings are certainly not connected at random.  Our $k$-shells are connected in a more complex manner, but seem to be described by simple rules.

\begin{figure}[htb] \centering
\includegraphics[totalheight=3.5in]{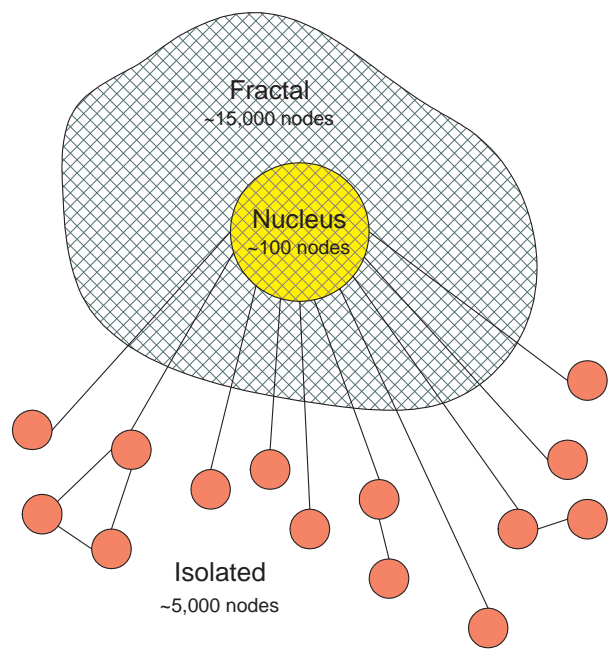}
\caption{\small{Medusa model of the AS network.}} \label{Meduza}
\end{figure}

\begin{figure}[htb] \centering
\includegraphics[totalheight=4.0in]{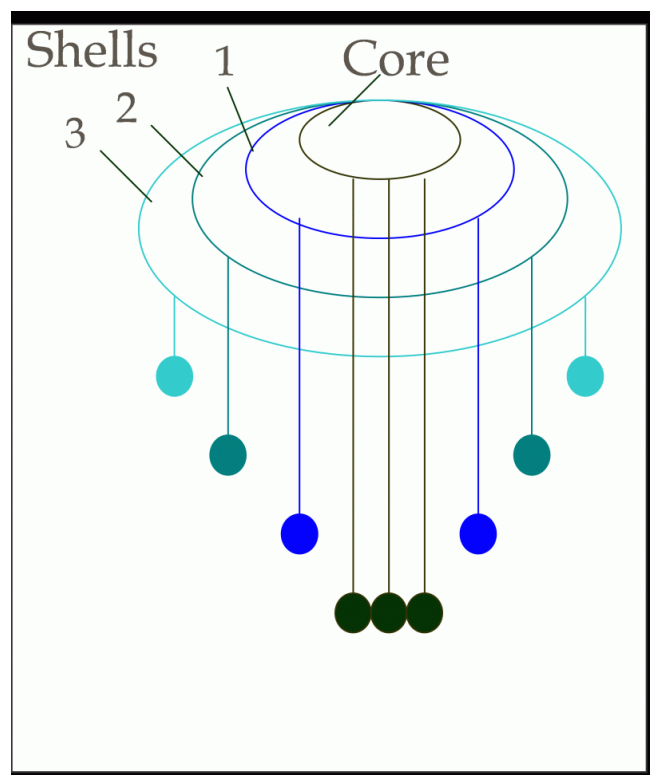}
\caption{\small{Jellyfish picture of the AS network. From \cite{TPSF01}}} \label{Jellyfish}
\end{figure}

We can visualize our model in a simple picture, shown as Figure \ref{Meduza}, which makes the differences between this view and the "Jellyfish" of Figure \ref{Jellyfish} apparent.  The core of the Medusa includes the most important nodes that are found in the core and the first ring of the Jellyfish's mantle.  The Jellyfish has relatively few rings around its core, while the Medusa's mantle is more extended and differentiated.  The tendrils hanging from the Jellyfish (leaf nodes) descend mostly from the core, but also from all the other rings, while all the tendrils of the Medusa are, by construction, attached to its nucleus.    

Proving that a fractal structure exists in the Medusa graph requires more than just exhibiting a power law for the sizes of the respective $k$-shells or $k$-cores.  We need to demonstrate that an appropriate renormalization scheme compresses ("rescales") the network in a simple way, preserving its structure.  Such a rescaling can be carried out as suggested by Song et al.\ \cite{SHM05} by collapsing clusters of nodes of a fixed diameter into pseudo-nodes of the renormalized graph.  The most natural way to assign nodes to boxes is to attempt (with heuristics) to find a covering using the minimum number of boxes of a given diameter, but Song et al. observe that the specific algorithm used seems to have little effect on the rescaling accomplished.  If there is a fractal structure preserved then the number of boxes or pseudonodes of diameter $l, N(l)$ should vary as $N(l) \propto l^{-d_f},$ where $d_f$ is a fractal dimension.  $d_f = 2$ is the expected result for percolation on infinite dimensional random networks \cite{CHA02}. (The argument for this is essentially the following:  above 6 dimensions (and in random E-R graphs), $d_f$ is 4 but we are measuring length along shortest paths, which in high dimensions behave like random walks, whose length goes as the square of a Euclidean dimension.  
Thus as a function of $l$, the diameter in terms of path length, we expect to see the number of boxes decrease as $l^{-2}$.)

\begin{figure}[htb] \centering
\includegraphics[totalheight=3.5in]{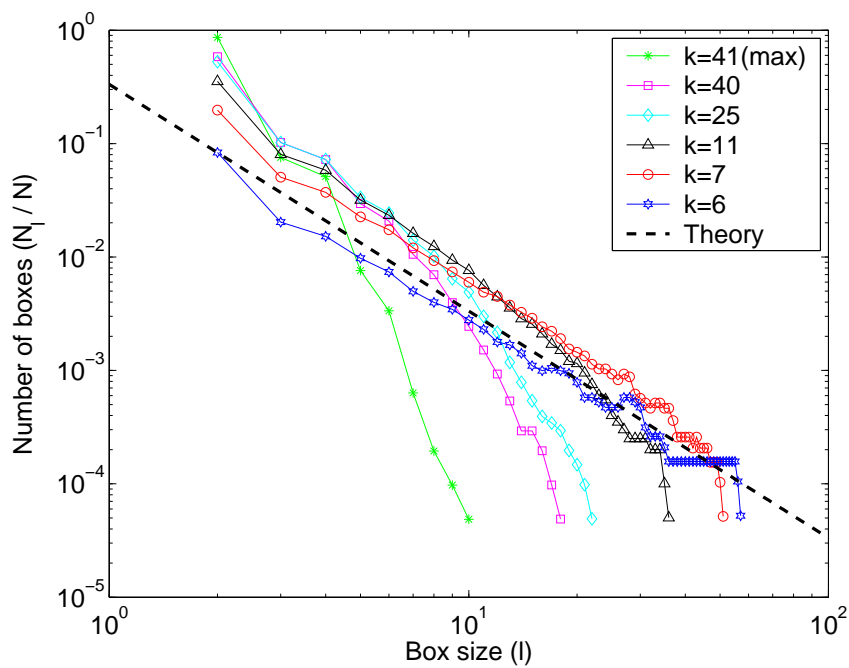}
\caption{\small{Results of renormalizing the connected components of crusts of the AS graph, using the "box cover" transformation, with  boxes constructed heuristically to ensure a near minimal covering. The "theory" line simply indicates a slope of -2, as argued in the text.}} 
\label{box_cover}
\end{figure}

\begin{figure}[htb] \centering
\includegraphics[totalheight=3.5in]{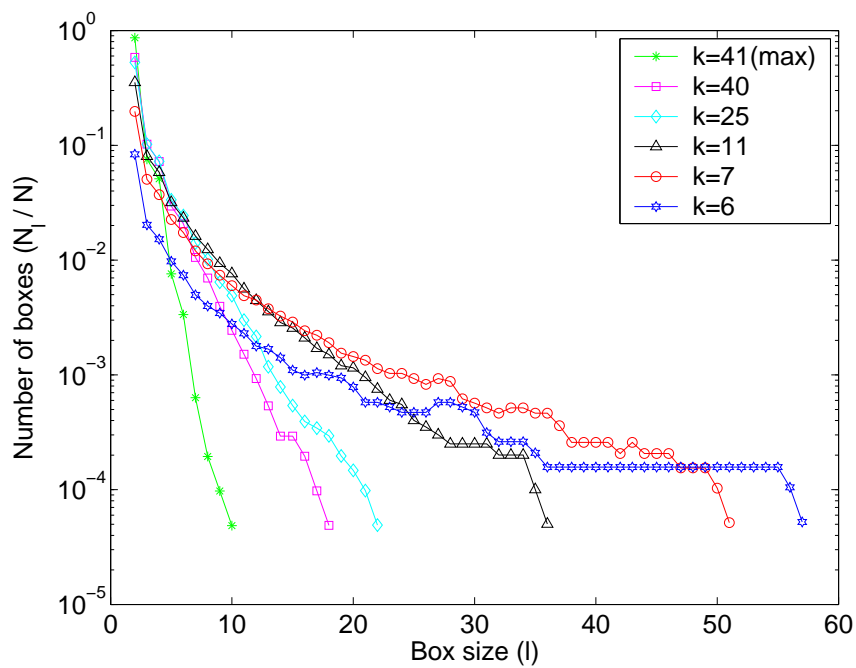}
\caption{\small{Data of Figure \ref{box_cover} plotted as a semilog plot, to test for exponential decay of the covering volume at larger distances. }} 
\label{box_coversemilog}
\end{figure}

We performed a "box cover" reduction on the AS graph and on each of its crusts, using a heuristic search method which should provide nearly the minimal number of boxes required to cover.  The nodes are sorted in decreasing order by initial degree.  Each box is centered on the node remaining uncovered which had the highest initial degree. Results shown in Figure \ref{box_cover}.  The dashed line drawn in Figure \ref{box_cover} indicates a slope of -2, or $d_f = 2$.  
Each curve in Fig \ref{box_cover} was derived by rescaling a different crust.  When the whole graph is rescaled ($k=41$), the volume vs box diameter curve does not follow the $l^{-2}$ slope.  Its collapse follows a much steeper slope.  An exponential decrease is to be expected in rescaling an infinite dimensional graph and a semilog plot of the same data (Figure \ref{box_coversemilog}) is consistent with this interpretation.   
The curves describing crusts 40 to 21 show a crossover behavior.  
The initial renormalization, using small boxes, shows the mean field slope, but the rescaling process ultimately steepens.  The crusts 21 and smaller do agree over most of the rescaling with the mean field prediction.  The biggest change in this characteristic comes when the core is removed, taking with it over half of the sites with highest degree.  A simple interpretation of Figure \ref{box_cover} is that it shows the AS graph to consist of two regimes -- a star-like "small world" regime (in which the nodes of highest degree connect to a large fraction of everything else) that causes the box-cover to decrease exponentially in volume, and a fractal inflow regime which characterizes the structure of the outer crusts.  The crossover point between the two regimes can be read off from Figures 
\ref{box_cover} and \ref{box_coversemilog}, and provides an correlation length within which the shallower crusts remain fractal.

\section{Conclusions and Future Work}

We have shown that the AS graph can be decomposed into elements that have distinct natural roles, using the technique of $k$-pruning, and capture these insights in a model, the Medusa model, which can be used to track changes in the Internet's composition as it evolves in time.  The nucleus or heart of the network which we identify is a robustly defined measure of what has in the past been called "Tier One" of the Internet -- the nodes which form a highly interconnected cluster with diameter 2, greater than 70 per cent interconnectivity, and high betweenness, suggesting that there must be very high bandwidth and capacity present as well.  Viewed as a graph, our Medusa structure is a novel combination of a fractal and a "small world" core. It is possible that a single process can be responsible for both aspects, since the fractal regime appears to be confined to within a critical correlation length of the leaves of the network.

Two key problems present attractive areas for future work.  The first is to use these more accurate measurements as a source of parameters for an improved generator of random models for the internet both now, and in future years.  

The second is to use the $k$-shell index to generalize the analysis of Internet redundancy and vulnerability to an intelligently directed attack that was initiated by Cohen et al. in \cite{CEAH01}.  Since the ASes in the nucleus of Internet prove to be the best-known 
carriers in many countries, an attack which disables ASes in descending order of their $k$-shell index would roughly correspond to an attack by reputation.  This might be a more plausible hacker scenario than one in which the attack proceeds in order of node degree, since node degree is not public knowledge.  However, to understand the vulnerability of Internet routing, described with the increased accuracy that this data permits, we must first discuss the differences between BGP routing and the much richer interconnection that is assumed in percolation analysis, and this we leave for a future paper.

\section{Acknowledgements}

The DIMES measurements and our analysis are parts of the EVERGROW European integrated project No. 1935, funded within the Sixth Framework's Future and Emerging Technologies division.  We also acknowledge support from the Israel Science Foundation, the Israel Internet Association, and the European NEST/PATHFINDER project DYSONET 012911.  Conversations with Sorin Solomon, Avishalom Shalitt, Alessandro Vespignani and Dietrich Stauffer 
are also gratefully acknowledged.  
S.K. would like to thank ICSI (at U.C. Berkeley) for its hospitality during summer 2005, when parts of this work were carried out.

%\begin{figure}[htb] \centering
%\includegraphics[totalheight=3.5in]{IntentionalandCrustsRealandRandom.eps}
%\caption{\small{Results of two types of targeted attacks on the AS graph in %reducing the size of its largest connected component.  The "random" graph %with which we compare was constructed by prescribing the degree %distribution to be that of the actual AS graph data, but with 10x as many %nodes.  Two strategies for the attack are "By Degree," which eliminates %nodes in order of their degree, largest first, and "By Shell," which %eliminates nodes by $k$-shell, starting with the largest k.}} 
%\label{Attacks}
%\end{figure}

\end{document}